\documentclass[useAMS,usegraphicx,usenatbib]{mn2e}
\usepackage{color}
\usepackage{epsf}
\usepackage{amssymb}

\def\be{\begin{equation}}
\def\ee{\end{equation}}

\def\Ngal{N_{\rm gal}}
\def\simle{\lesssim}

\def\RMS{\mbox{RMS}}

\title{No evidence for the cold spot in the NVSS radio survey}

\author[Kendrick M. Smith and Dragan Huterer]
{Kendrick M. Smith$^{1}$ and Dragan Huterer\vspace{0.2cm}$^{2}$\\
$^{1}$ Institute of Astronomy, Cambridge University, Madingley Road, Cambridge CB3 0HA, UK\\
$^{2}$ Department of Physics, University of Michigan, 450 Church St,  Ann Arbor MI 48109, USA}
\begin{document}


\pubyear{2007}

\maketitle

\label{firstpage}

\begin{abstract}
We revisit recent claims that there is a ``cold spot'' in both number counts
and brightness of radio sources in the NVSS survey, with location coincident with
the previously detected cold spot in WMAP.  Such matching cold spots would be
difficult if not impossible to explain in the standard $\Lambda$CDM
cosmological model.  Contrary to the claim, we find no significant evidence
for the radio cold spot, after including systematic effects in NVSS, and
carefully accounting for the effect of {\em a posteriori} choices when assessing
statistical significance.
\end{abstract}


\section{Introduction}

Cosmic microwave background (CMB) maps have been studied in detail during the
last few years. These studies have been motivated by the remarkable full-sky
high-resolution maps obtained by WMAP \citep{Bennett2003,Spergel:2006}, and
led to a variety of interesting and unexpected findings.  Notably, various
anomalies have been claimed pertaining to the alignment of largest modes in
the CMB
\citep{deOliveira2004,Hajian2003,SS2004,TOH,Schwarz2004,Land2004a,Land2005a,lowl2},
the missing power on large angular scales \citep{Spergel2003,wmap123}, and the
asymmetries in the distribution of power
\citep{Eriksen_asym,Bernui:2006ft,Hajian:2007pi}.  In the future, temperature
maps obtained by the Planck experiment, and large-scale polarization
information \citep{Dvorkin} may be key to determining the nature of the
large-scale anomalies.  For a review of the anomalies and attempts to explain
them, see \citet{Huterer_review}.

Recently, a paper by \citet{Rudnick} attracted particular attention, as it
claimed to have detected a 'cold spot'' --- a drop in the source density {\it
  and} brightness in the NVSS survey.  This
claim would be relatively unremarkable, if it were not for the fact that a
previously reported, anomalously cold spot in the WMAP microwave signal
\citep{Vielva2004,Cruz2005,Cruz2006,Cruz2007,Cayon2005} apparently 
lies at roughly the same location.

This claim, if verified to be true at high statistical significance, would
represent a major result, and would be difficult or impossible to explain in
the standard cosmological model.  One interpretation, proposed in
\citet{Inoue_Silk} and \citet{Rudnick}, is the existence of a large ($\gtrsim
100$ Mpc) void at $z\sim 1$, which gives rise to an NVSS underdensity directly,
and gives rise to the WMAP cold spot via the nonlinear ISW effect.  However,
the probability of forming such a large void in $\Lambda$CDM cosmology is
negligibly small.

Here we reexamine these claims using our own analysis procedure, carefully
including known systematic and statistical properties of NVSS
(declination-dependent ``striping'' and galaxy-galaxy correlations; see
\S\ref{ssec:case1}), and marginalizing {\em a posteriori}
choices when assessing statistical significance.  We will argue that there is
no statistically significant evidence for either a dip in NVSS number counts
or median source flux in the WMAP cold spot.
We will see that it is possible to construct statistics containing {\em a posteriori} choices
(e.g., the location and radius of a ``sub-disc'' of the cold spot)
which might appear to support an underdense region, but the statistical signifiance
goes away when these choices are properly marginalized.
Furthermore, we will show that by making different {\em a posteriori} choices, we could
find evidence for an {\em overdense} region with the same statistical significance
as an underdense region.

The paper is organized as follows.  In \S\ref{sec:data} we describe the NVSS
data and selection cuts that we consider.  In \S\ref{sec:galaxy_counts} we
perform statistical tests using the number density distribution of NVSS
sources, and in \S\ref{sec:flux_maps} we do the same for the flux
distribution.  In \S\ref{sec:cuts} we study the dependence of our results on
the selection cuts that are applied to the NVSS catalog prior to the analysis.
Our main result, showing significance of anomalous number counts or source
fluxes in the WMAP cold spot, using several different statistics and with a
range of possible selection cuts in the NVSS catalog, is shown in Table~\ref{tab:cuts}.
We conclude in \S\ref{sec:conclude}.

\section{Data} 
\label{sec:data}

\begin{figure}
\centerline{\epsfxsize=3.2truein\epsffile{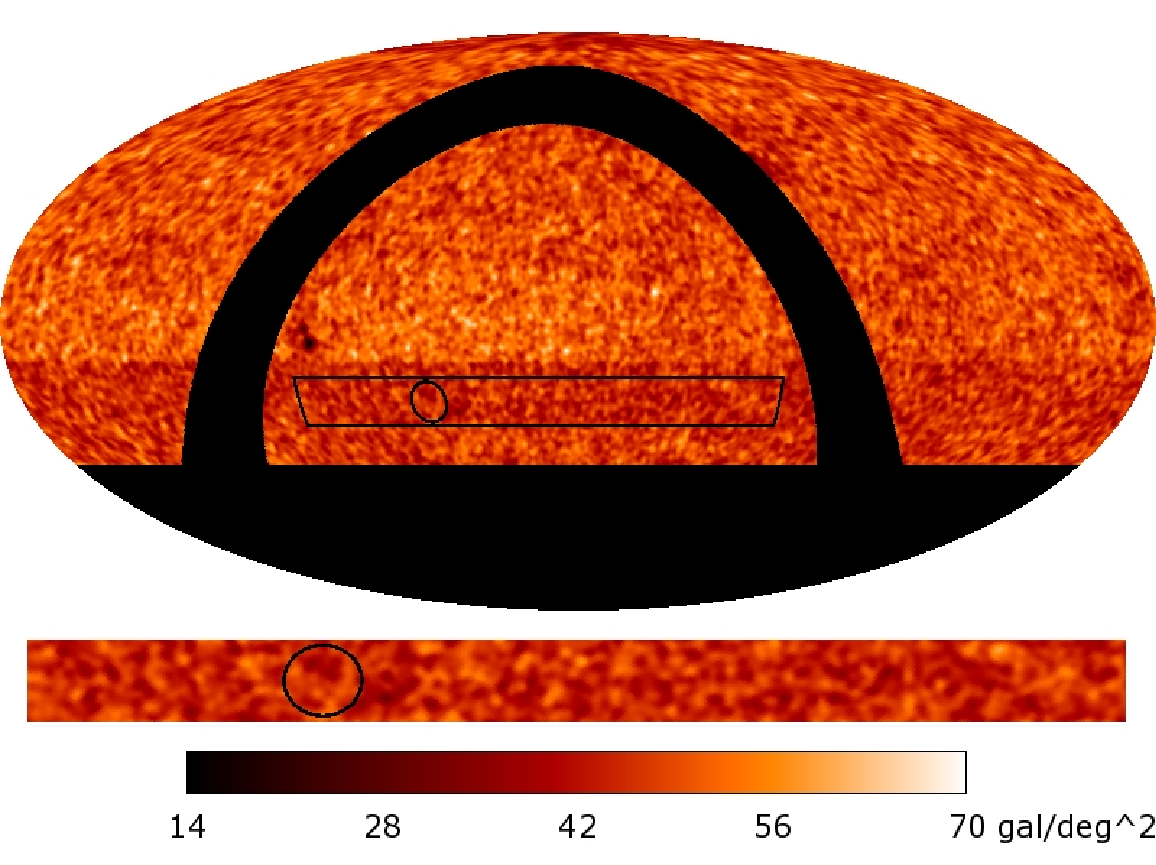}}
\caption{Galaxy count maps for the NVSS survey, smoothed to $1^\circ$ radius and
plotted in equatorial coordinates, with default data cuts as described in \S\ref{sec:data}.
The full sky counts (top panel) show declination-dependent variations in
completeness level; the WMAP cold spot (shown as a circle in both panels) is entirely
contained within the underdense ``stripe'' at declination $\delta\simle -10^\circ$.
If we zoom in on a box at the same declination as the cold spot, then the WMAP cold spot
does not look anomalous by eye (bottom panel; shown as a rectangular region in the top panel).}
\label{fig:galaxy_count_maps}
\end{figure}

The NRAO VLA Sky survey (NVSS) is a 1.4 GHz continuum survey, covering 82\% of
the sky, with a source catalog containing over $1.8\times 10^6$ sources that
is 50\% complete at 2.5 mJy.  Away from the galactic plane, almost all of the
sources are extragalactic: quasars, or AGN-powered or star-forming galaxies.
The NVSS catalog covers a wide range of redshifts (the median redshift is
$z\approx 0.9$), but dividing the catalog into redshift bins is not possible
because per-source redshifts are not measured.
However, in \S\ref{sec:cuts} we will explore the effect of dividing the catalog into flux bins.

When making Healpix \citep{Gorski:2004by} maps from the NVSS catalog, we mask
pixels near the galactic plane ($|b| < 10^\circ$) or the boundary of the
survey (declination $\delta < -37^\circ$).  Our ``default cuts'' will consist of this
pixel mask, plus dropping all sources which are flagged in the NVSS catalog as having complex
structure.  (These are mainly galactic sources; if they are included in the maps then spurious
features can be seen by eye at low galactic latitude.)
We will also consider other choices of cuts in \S\ref{sec:cuts}.

In Fig.~\ref{fig:galaxy_count_maps}, we show a full-sky map and a zoomed-in
region near the WMAP cold spot, with default selection cuts and smoothed to
$1^\circ$ resolution.  We have shown the full-sky map in equatorial
coordinates to highlight a known systematic effect in NVSS
\citep{Blake:2001bg}: the presence of declination-dependent variations in
completeness level, most notably the underdense ``stripe'' at $\delta\simle
-10^\circ$.  Because the WMAP cold spot is inside the stripe, modeling these
variations will play an important role in the analysis, as we will see in
detail below.

\section{Galaxy count analysis}
\label{sec:galaxy_counts}

In \cite{Cruz2005}, the WMAP cold spot is given as a circular region with
center $P_0$ at $(l,b)=(209^\circ, -57^\circ)$ in galactic
coordinates, and radius $r_0=5^\circ$.
However, in \cite{Rudnick} the most anomalously underdense circular region in NVSS
is quoted as having center $P_0'$ at $(l,b)=(207.03,-54.85)$ and radius $r_0'=1^\circ$.
How does this mismatch between the WMAP cold spot and the NVSS underdense region affect
the analysis?

The total statistical significance of the WMAP cold spot is only $\approx 3\sigma$.
Presumably, at this significance,
the best-fit center and radius have non-negligible statistical errors,
and any nontrivial shape or substructure of the cold spot is not resolved.

For this reason, it seems reasonable to look for an NVSS underdensity $(P_0',r_0')$ which need not be equal 
to the WMAP cold spot $(P_0,r_0)$.
However, this makes statistical significance more difficult to assess: for a correct treatment, the parameters 
$(P_0',r_0')$ must be treated as {\em a posteriori} choices.
Alternately, one could simply ask whether the region $(P_0,r_0)$ is underdense in NVSS.  In this case there are no
{\em a posteriori} choices (we are simply counting NVSS galaxies using the best-fit cold spot parameters from WMAP data alone)
and computing statistical significance is straightforward.

Our perspective is that either of these analyses is valid; in the next three subsections we consider the
following possibilities for a disc-shaped NVSS underdensity with center $P$ and radius $r$:

\begin{enumerate}
\item $P=P_0, r=r_0$: The NVSS underdensity has the same center and radius as the WMAP
  cold spot.

\item $P=P_0, r\ne r_0$: The NVSS underdensity has the same center as the WMAP cold spot
  but its radius is different; then we must assign statistical significance in
  a way which incorporates the {\em a posteriori} choice of radius.

\item $P\subseteq P_0, r\ne r_0$: The NVSS underdensity lies wholly within the WMAP
  cold spot but both its center and its radius are different; then we
  must incorporate the {\em a posteriori} choice of both radius and location.
\end{enumerate}

\subsection{Case 1: Fixed center, fixed radius}
\label{ssec:case1}

This case ($P=P_0,r=r_0$) corresponds to the simplest possible question: if we
count the total number of galaxies in the WMAP cold spot, do we get an
anomalous value?  We introduce the ratio statistic, \be
\frac{\Ngal(P_0,r_0)}{\langle\Ngal(P_0,r_0)\rangle} \label{eq:ratio} \ee and
ask whether it differs from 1.0 with statistical significance, where the
numerator $\Ngal(P_0,r_0)$ denotes the number of galaxies in the WMAP cold
spot $(P_0,r_0)$ and the denominator is its expectation value.

Two issues arise here: first, how should the denominator
$\langle\Ngal(P_0,r_0)\rangle$ be computed?  The simplest prescription would
be to assume that the expected number density per unit area is equal to the
full-sky NVSS average:
\be
\langle \Ngal(P,r) \rangle = \pi r^2 \langle n \rangle_{\rm full-sky}    \label{eq:nexp1}
\ee
where $ \langle n \rangle_{\rm full-sky}$ is the mean number density per unit area
on the full NVSS sky.  This
simple estimate for $\langle\Ngal\rangle$ is not satisfactory because it does
not account for the underdense stripe (see Fig.~\ref{fig:galaxy_count_maps}).
Therefore, we also consider an improved prescription based on a simple stripe
model.  We assume that the expected number of galaxies in each Healpix pixel
$p$ is equal to the average taken over unmasked pixels $p'$ at the same
declination: \be \langle \Ngal(P,r) \rangle = \sum_{p \in(P,r)} \langle
\Ngal(p') \rangle_{p'\sim p}
\label{eq:nexp2}
\ee 
where the sum runs over pixels $p$ in the disc $(P,r)$, and $\langle
\Ngal(p') \rangle_{p'\sim p}$ denotes the average galaxy count taken in pixels
$p'$ at the same declination as $p$.

The second issue when studying the ratio statistic in Eq.~(\ref{eq:ratio}) is
how error bars should be assigned.  Here, the simplest prescription would be
to assume Poisson statistics: we take the uncertainty in the numerator to be
given by \be \Delta \Ngal(P,r) =
\langle\Ngal(P,r)\rangle^{1/2} \label{eq:delta1} \ee This simple prescription
for $(\Delta\Ngal)$ underestimates the error bars because it assumes that the
NVSS galaxies are pure shot noise, i.e.  galaxy-galaxy correlations are
ignored.\footnote{In principle, this could be remedied by estimating the
  galaxy power spectrum and including it in Monte Carlo simulations, although
  this may be difficult in practice, due to the presence of long-wavelength
  instrumental power in NVSS at low flux levels \citep{Smith:2007rg,Ho:2008bz} 
  which may
  not be accurately modeled by an isotropic Gaussian field.  In
  Eq.~(\ref{eq:delta2}), we have taken a simpler approach by averaging over
  choices of disc center $P'$ in the real NVSS data, rather than averaging
  over Monte Carlo simulations.}  Therefore, we also consider an improved
prescription: we estimate $(\Delta\Ngal)$ directly from the data by taking the
RMS fluctuation over alternate choices $P'$ of ring center which lie at the
same declination as $P$: \be \frac{\Delta
  \Ngal(P,r)}{\langle\Ngal(P,r)\rangle^{1/2}} = \RMS_{P'\sim P} \left(
  \frac{\Ngal(P',r) -
    \langle\Ngal(P',r)\rangle}{\langle\Ngal(P',r)\rangle^{1/2}} \right)
\label{eq:delta2} 
\ee 
where the notation $\RMS_{P'\sim P}(\cdot)$ denotes the
RMS fluctuation taken over choices of center $P'$ with the same declination as
$P$.  \footnote{As a minor point, in Eq.~(\ref{eq:delta2}) we have normalized
  each value of $\Ngal$ by its Poisson error $\langle\Ngal\rangle^{1/2}$
  before taking the RMS; this slightly improves the estimate of
  $(\Delta\Ngal)$ by accounting for variations in $\langle\Ngal\rangle$ due to
  striping and the pixel mask.}

With default cuts (\S\ref{sec:data}), we find the following results.  If we
use full-sky averaging (Eq.~(\ref{eq:nexp1})) and Poisson errors
(Eq.~(\ref{eq:delta1})), then the WMAP cold spot appears to be underdense in
NVSS sources at 3.1$\sigma$.  However this is simply an artifact of using the full-sky
average galaxy density when computing $\langle\Ngal\rangle$; in fact, the WMAP
cold spot is contained within the underdense NVSS stripe
(Fig.~\ref{fig:galaxy_count_maps}).  If we improve the estimate of
$\langle\Ngal\rangle$ by using isolatitude averaging (Eq.~(\ref{eq:nexp2}))
then the cold spot appears {\em overdense} at 1.1$\sigma$.  This is already
not statistically significant, but if we improve the estimate of
$\Delta\Ngal(P,r)$ using Eq.~(\ref{eq:delta2}), then the overdensity drops to
0.8$\sigma$.  We conclude that the WMAP cold spot, taken as a whole, is not
underdense or overdense in NVSS sources, but modeling the NVSS ``stripe''
plays a crucial role in the analysis.

We note that declination-dependent striping is unlikely to affect the analysis of \cite{Rudnick}, which is 
restricted to small regions of sky near the cold spot.  The discussion here is intended to motivate
expressions such $(\Delta\Ngal)$ above (Eq.~(\ref{eq:nexp2}), and related quantities later in the paper,
which are defined in a way which is robust to striping.

\subsection{Case 2: Fixed center, floating radius}
\label{ssec:case2}

We next consider the possibility of an underdense region with the same center
$P=P_0$ as the WMAP cold spot but arbitrary radius $r<r_0$.  We again define a
ratio statistic \be \frac{\Ngal(P_0,r)}{\langle\Ngal(P_0,r)\rangle} \ee and
compute the expectation value $\langle\Ngal(P_0,r)\rangle$ and RMS deviation
$\Delta\Ngal(P_0,r)$ following the discussion in the preceding subsection.

\begin{figure}
\centerline{\epsfxsize=3.2truein\epsffile[18 250 592 718]{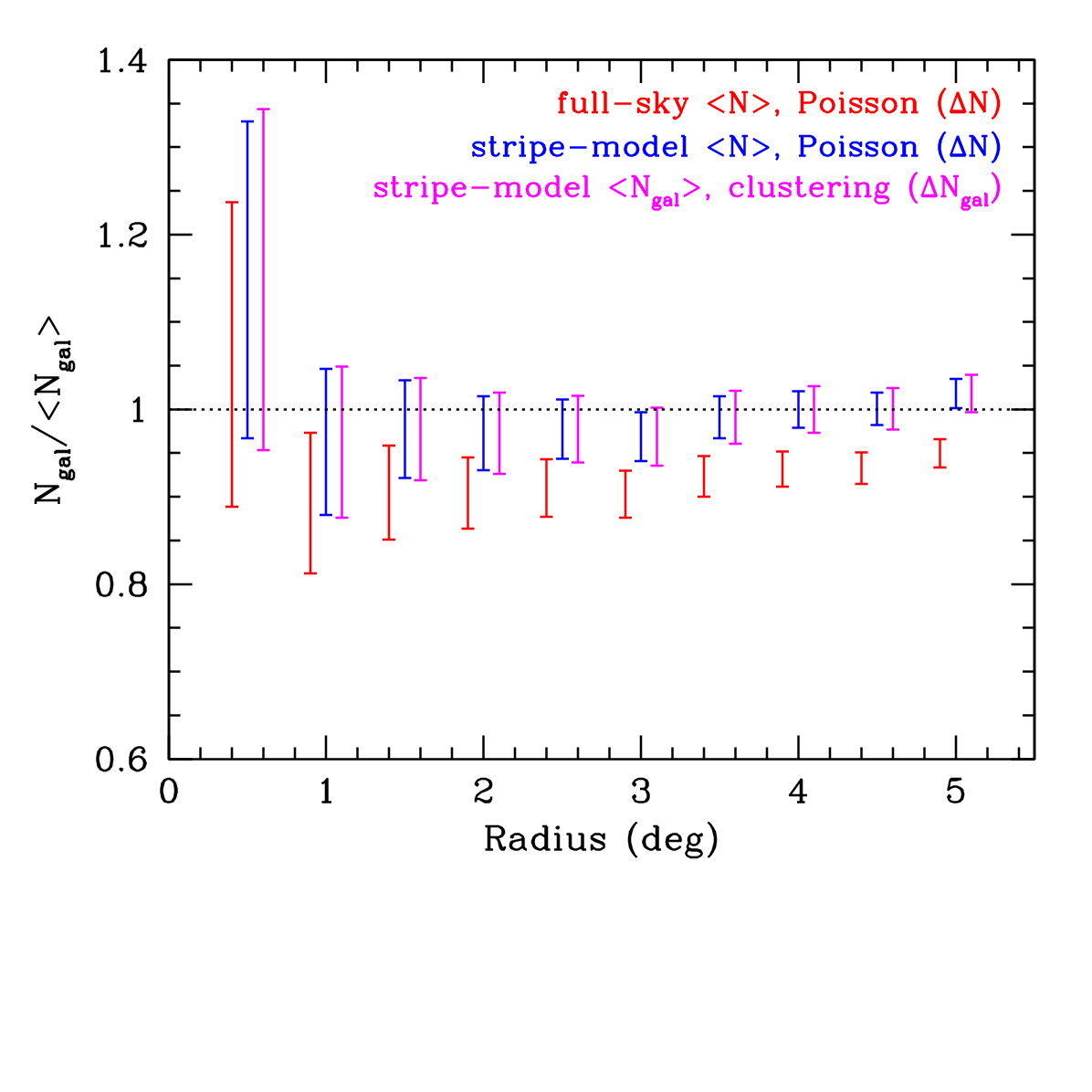}}
\caption{Galaxy counts in circles of varying radii centered at the WMAP cold
spot location, relative to expected counts as in Eq.~(\ref{eq:ratio}).  From
left to right, the three sets of error bars (all 68\% C.L.) represent increasingly accurate
analysis methods.  The left errorbars assume full-sky $\langle\Ngal\rangle$ and
Poisson $(\Delta\Ngal)$ (Eqs.~(\ref{eq:nexp1}),~(\ref{eq:delta1})).  The middle
and right errorbars incorporate declination striping in $\langle\Ngal\rangle$
and galaxy-galaxy clustering in $(\Delta\Ngal)$ respectively
(Eqs.~(\ref{eq:nexp2}),~(\ref{eq:delta2})).  It is seen that if the NVSS
underdense stripe is not modeled, then the WMAP cold spot appears to be anomalously
underdense, but the statistical significance is lost when the
stripe is included in the analysis.}
\label{fig:galcount_analysis}
\end{figure}

Results using this statistic are shown (with default cuts) in
Fig.~\ref{fig:galcount_analysis}.  The rightmost error bars represent our most
accurate ways of computing $\langle\Ngal\rangle$ and $(\Delta\Ngal)$
(Eqs.~(\ref{eq:nexp2}),~(\ref{eq:delta2})); we find that all points are within
$1\sigma$ of the expected level, i.e. no evidence for an underdensity is seen.
(We note that if say, one value of $r$ gave an anomalous value, then we would
have to incorporate the {\em a posteriori} choice of $r$ when assessing
significance; we revisit this issue in \S\ref{sec:cuts}.)  The left and middle
error bars represent less accurate ways of computing $\langle\Ngal\rangle$ and
$(\Delta\Ngal)$ (Eqs.~(\ref{eq:nexp1}),~(\ref{eq:delta1})) and are shown for
comparison.

\subsection{Case 3: Floating center, floating radius}
\label{ssec:case3}

The result of the preceding subsection appears to contradict Fig.~5 in
\cite{Rudnick}, where a statistically significant underdense disc of radius
$r_0'=1^\circ$ is seen.  However, this figure has been constructed taking the
disc center $P_0'$ to be the point $(l,b)=(207.03,-54.85)$ rather than the
center $P_0$ of the WMAP cold spot which is at $(l,b)=(209, -57)$.

If the choice of $(P_0',r_0')$ had an {\em a priori} motivation, then we would
find, using an analysis similar to \S\ref{ssec:case1}, a 2.0$\sigma$
underdensity, with our default cuts.  (There are other underdense ``subdiscs''
as well, e.g. we find that the subdisc centered at $(l,b)=(206.82, -56.4)$
with radius $2.5^\circ$ is underdense at 3.06$\sigma$ if {\em a posteriori} choices
are ignored.)  However, we see no {\em a priori} motivation for making such
choices of $(P_0',r_0')$, and we must therefore incorporate the effect of the
{\em a posteriori} choice when calculating statistical significance.

To assess significance fairly, we incorporate the effect of the choice of
$(P_0',r_0')$ as follows.  Formally, given a disc $(P,r)$, define the ``naive
number of sigmas'' of its worst underdense subdisc by: \be N_\sigma(P,r) =
\min_{(P',r')\subset (P,r)} \left( \frac{\Ngal(P',r') -
    \langle\Ngal(P',r')\rangle}{\Delta\Ngal(P',r')}
\right) \label{eq:defnsigma} \ee where the notation $\min_{(P',r')\subset
  (P,r)}(\cdot)$ means that the quantity in parentheses is minimized over
discs $(P',r')$ which are entirely contained in $(P,r)$, with minimum radius
$r'_{\rm min}=1^\circ$.
 In this notation, the existence
of the ``subdisc'' $(P_0',r_0')$ from the preceding paragraph can be rephrased as the
statement that $N_\sigma(P_0,r_0)=-3.06$, where $(P_0,r_0)$ are the cold spot
center and radius respectively.

To assess whether this value is anomalous, we evaluate the same statistic
(``number of sigmas of the worst underdense subdisc'') in an ensemble of disc-shaped
regions with the same size and declination as the WMAP cold spot.
This way of assessing significance accounts for
both the {\em a posteriori} choice of $(P_0',r_0')$ and declination-dependent
striping.
More precisely, we compute $N_\sigma(P,r_0)$ for an ensemble of alternate choices
of disc center $P$ with the same declination as the WMAP cold spot center $P_0$, and
with radius $r_0=5^\circ$.
We find that this ensemble of values has mean $\langle N_\sigma\rangle=-2.65$ 
and RMS error $(\Delta N_\sigma)=0.45$.  Therefore, the cold spot (with $N_\sigma=-3.06$)
is typical among discs with the same radius and declination, and the
``subdisc'' described above does not have statistical significance after the effect of 
{\em a posteriori} choices is taken into account.

\section{Flux analysis}
\label{sec:flux_maps}

\begin{figure}
\centerline{\epsfxsize=3.2truein\epsffile{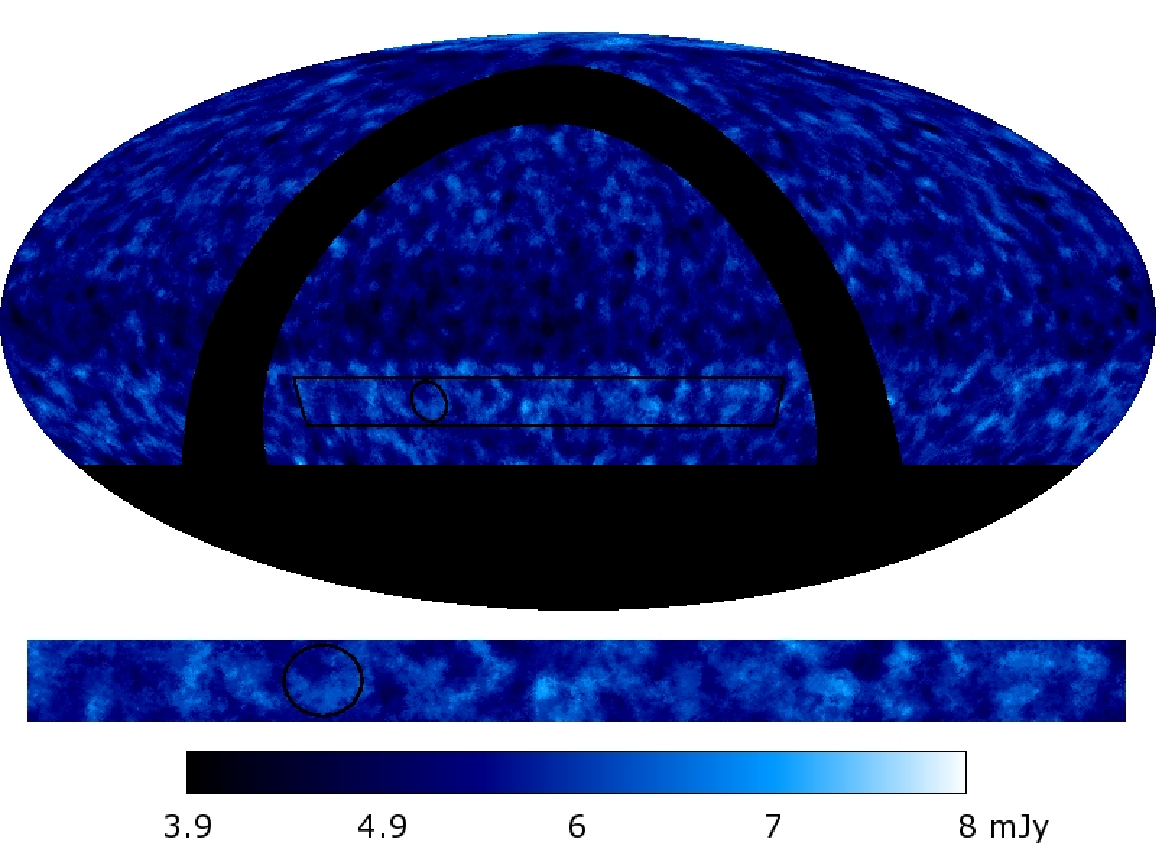}}
\caption{NVSS flux maps, smoothed by $2^\circ$ median filtering as described
in \S\ref{sec:flux_maps}, shown with default cuts in equatorial coordinates.
As in the galaxy count case (Fig.~\ref{fig:galaxy_count_maps}), the full-sky
map (top panel) shows declination-dependent variations in median flux, and the
WMAP cold spot (shown as a circle) is not anomalous by eye when viewed in a
``box'' at the same declination (bottom panel).}
\label{fig:flux_maps}
\end{figure}

In addition to number counts, the median NVSS brightness was also reported to
be low near the WMAP cold spot in \cite{Rudnick}.  Since brightness is roughly
proportional to (source counts) $\times$ (source flux), and we have already
analyzed the source counts, our perspective is that it is better to separate
the two issues and next ask whether median source fluxes in NVSS are anomalous
in the WMAP cold spot.  Considering source counts and fluxes separately,
rather than using brightness maps, has two additional advantages.  First, it allows
the analysis to proceed from the NVSS source catalog, thus avoiding
instrumental complexities associated with working with the NVSS images, which
have been incorporated by the NVSS team when constructing the source catalog.
Second, it avoids introducing more {\em a posteriori} choices which must be
marginalized (e.g. in \cite{Rudnick}, the brightness maps are convolved with
an 800 arcsec filter to obtain a continuous field, which is then
median-filtered in sliding boxes with side length 3.4$^\circ$; a ``dip'' is
then observed at a point other than the WMAP cold spot center.)

We would also like to emphasize that, if the purpose of this analysis is to find voids,
then there is no {\em a priori} motivation for considering either brightness or source fluxes;
the best-motivated statistics would be based on number counts alone.
Nevertheless, in this section, we will briefly NVSS source fluxes
in the WMAP cold spot.
Our median flux analysis will be analogous to the galaxy count case from the preceding
section; we summarize our methodology and results here.

In Fig.~\ref{fig:flux_maps}, we show a flux map obtained by taking the median
flux of all NVSS sources within $2^\circ$ of each pixel.  This median-based
smoothing procedure was used because the NVSS flux distribution contains far
outliers; if the mean were used instead of the median, then the map would be
dominated by a small number of rare bright sources.  Note that
declination-dependent striping is seen in the flux map, as seen previously for
number counts (Fig.~\ref{fig:galaxy_count_maps}).

Given disc center $P$ and radius $r$, we define $\mu(P,r)$ to be the
median flux of all NVSS galaxies contained in the disc $(P,r)$, and consider ratio 
statistics of the form:
\be
\frac{\mu(P,r)}{\langle\mu(P,r)\rangle}  \label{eq:ratio2}
\ee
We estimate the expected median flux $\langle\mu(P,r)\rangle$ directly from the
data by averaging over alternate choices of center $P'$ with the same
declination as $P$:
\be
\langle \mu(P,r) \rangle = \langle \mu(P',r) \rangle_{P'\sim P}   \label{eq:muexp}
\ee
We also estimate the error $(\Delta\mu(P,r))$ from the data in an analogous way\footnote{Note
  that we have taken the factor $\langle\Ngal\rangle^{1/2}$ inside the RMS average,
  to improve the estimate of $(\Delta\mu)$ by accounting for the scaling $(\Delta\mu \propto \Ngal^{-1/2})$
  expected due to variations in number density alone.}
\be
\Delta\mu(P,r) =
\frac{\RMS_{P'\sim P} \langle\Ngal(P',r)\rangle^{1/2} \big(\mu(P',r)-
  \langle\mu(P',r)\rangle\big)}{\langle\Ngal(P,r)\rangle^{1/2}}
\label{eq:deltamu}
\ee

We then consider three cases for the disc $(P,r)$, as in \S{\ref{ssec:case1}}-\S{\ref{ssec:case3}}.
\begin{enumerate}
\item Fixed center, fixed radius ($P=P_0, r=r_0$): In this case, we simply
  evaluate the ratio statistic in Eq.~(\ref{eq:ratio2}), taking $(P,r)$ to be the WMAP cold
  spot center and radius $(P_0,r_0)$.  We find that the ratio exceeds 1.0 by 0.6$\sigma$,
  i.e. the median flux of all NVSS galaxies in the WMAP cold spot is not
  anomalous.
\item Fixed center, floating radius ($P=P_0, r<r_0$): In this case, we compute
  the ratio statistic in Eq.~(\ref{eq:ratio2}) for a variety of radii centered
  at the cold spot center $P_0$ (in analogy with
  Fig.~\ref{fig:galcount_analysis}).  We find that all values are within
  1.2$\sigma$ of 1.0, i.e. no anomalous value of the median flux is found.
\item Floating center, floating radius ($P\subseteq P_0, r < r_0$): In this
  case, we choose a subdisc $(P,r)$ of the WMAP cold spot $(P_0,r_0)$ which
  appears to give the most anomalous value of the ratio statistic in
  Eq.~(\ref{eq:ratio2}).  If we look for an anomalously low median flux, then
  we find that the subdisc with center $(l,b)=(206.81, -54.75)$ and radius $1^\circ$
  appears to be low at 2.2$\sigma$, if the {\em a posteriori} choice of
  $(P,r)$ is temporarily ignored.  To assess whether this value is really
  anomalous, we proceed in parallel with the number count analysis in
  \S\ref{ssec:case3}: we evaluate the same statistic (naive ``number of
  sigmas'' $N_\sigma$ of the most anomalous subdisc'') over an ensemble of
  regions with the same size and declination as the cold spot.  We get
  $N_\sigma=(2.2\pm 0.3)$ in this ensemble, i.e. the WMAP cold spot (with
  $N_\sigma=2.2$) is a typical member of this ensemble, and the low flux in
  the aforementioned subdisc has no statistical significance.
\end{enumerate}

\section{Alternate choices of cuts}
\label{sec:cuts}

\begin{table*}
\begin{tabular}{|c|c|c|c|c|c|c|}
\hline
  & \multicolumn{3}{c}{NVSS galaxy counts analysis} &
             \multicolumn{3}{c}{NVSS median flux analysis}  \\
\hline
Flux range    & WMAP center      & WMAP center,   & Any center,  & WMAP center      & WMAP center, & Any center, \\
    & {\em and} radius & any radius     & any radius   & {\em and} radius & any radius   & any radius  \\
\hline\hline
$S \le 3$ mJy & $-0.5\sigma$ ($-0.5\sigma$) & $-0.8\sigma$ ($-0.7\sigma$) & --- ($-0.0\sigma$) & $0.1\sigma$ ($0.2\sigma$) & $0.2\sigma$ ($0.2\sigma$) & --- (---) \\
$3 \le S \le 5$ mJy & $0.4\sigma$ ($0.2\sigma$) & --- (---) & --- (---) & $1.1\sigma$ ($0.9\sigma$) & $0.3\sigma$ (---) & $2.5\sigma$ ($-2.9\sigma$) \\
$5 \le S \le 12$ mJy & $2.6\sigma$ ($2.3\sigma$) & $2.0\sigma$ ($1.6\sigma$) & $1.7\sigma$ ($1.7\sigma$) & $0.4\sigma$ ($0.1\sigma$) & $-0.9\sigma$ ($-0.9\sigma$) & $1.4\sigma$ ($1.3\sigma$) \\
$S \ge 12$ mJy & $-0.4\sigma$ ($-0.6\sigma$) & --- ($-0.2\sigma$) & $-1.4\sigma$ ($-1.6\sigma$) & $1.3\sigma$ ($0.5\sigma$) & $-1.4\sigma$ ($-1.2\sigma$) & $-1.6\sigma$ ($-2.2\sigma$) \\
\hline
$S \le 5$ mJy & $-0.2\sigma$ ($-0.3\sigma$) & $-0.5\sigma$ ($-0.6\sigma$) & --- ($-0.1\sigma$) & $0.8\sigma$ ($0.8\sigma$) & $0.1\sigma$ ($0.1\sigma$) & $0.3\sigma$ ($0.2\sigma$) \\
$S \ge 5$ mJy & $1.4\sigma$ ($1.0\sigma$) & $0.6\sigma$ ($-0.2\sigma$) & $-1.8\sigma$ ($-2.5\sigma$) & $-1.7\sigma$ ($-1.7\sigma$) & $-1.1\sigma$ ($-1.0\sigma$) & $-1.6\sigma$ ($-2.0\sigma$) \\
\hline
Arbitrary $S$ & $0.9\sigma$ ($0.5\sigma$) & $-0.1\sigma$ ($-0.6\sigma$) & $-0.9\sigma$ ($-1.2\sigma$) & $0.7\sigma$ ($0.5\sigma$) & $0.4\sigma$ ($0.2\sigma$) & $-0.6\sigma$ ($-0.8\sigma$) \\
\hline
\end{tabular}
\caption{Statistical significance of anomalous NVSS number counts or median flux in the WMAP cold spot, for different NVSS flux ranges,
and with complex sources either dropped (unparenthesized values) or retained (parenthesized) in the analysis.
The six columns correspond to the different number count and median flux analyses that we have considered in \S\ref{sec:galaxy_counts}--\S\ref{sec:flux_maps}.
As described in \S\ref{sec:cuts}, each entry in the table is either the statistical significance of a region with high source density/flux (positive sign), 
or low source density/flux (negative sign), depending on which has higher significance.
An entry is marked `---' if we do not find any subdisc (either overdense or underdense) which is more anomolous than expected from statistics.
(This is assessed by comparing to an ensemble of regions with the same size and declination as the cold spot, as explained in detail in \S\ref{sec:cuts}.)
As described in the text, the NVSS source density is roughly 46 deg$^{-2}$, the flux ranges chosen in the table roughly divide the catalog into quartiles, and 
the disks used in the analysis have radii in the range $1 \le r \le 5$ deg.  Thus the disks used to construct the table contain between $\approx$36 and $\approx$3600
sources.}
\label{tab:cuts}
\end{table*}

We have now performed an exhaustive analysis of NVSS number density
(\S\ref{sec:galaxy_counts}) and median flux (\S\ref{sec:flux_maps}) in
subspots of the WMAP cold spot, with three cases depending on whether the
subspot location and radius are determined {\em a priori} or {\em a
  posteriori}, for a total of six analyses in all.  This has been done using
our ``default cuts'' from \S\ref{sec:data}: we drop NVSS sources flagged as
having complex structure to be conservative, but do not impose flux cuts, in
order to avoid making an {\em a posteriori} choice of flux range.

One possible loophole remains: in Fig.~5 in \cite{Rudnick}, the statistical
significance appears to be much higher if only sources with flux $S\ge 5$ mJy
are retained.  In this section, we consider the general question: can we get a
statistically significant result if we use selection cuts other than our default choice?

We divide the NVSS catalog into four flux bins, with delimiting values given
by $\{3, 5, 12\}$ mJy.
We also consider either dropping or retaining sources flagged
as having complex structure in the NVSS catalog.  For reference, the source
density of NVSS is 46 deg$^{-2}$ with complex sources dropped, or
49 deg$^{-2}$ with complex sources retained.  
The delimiting values for our four flux bins were chosen to further divide the
catalog into quartiles.

In Table~\ref{tab:cuts}, we summarize the results of repeating the six analyses
considered in this paper with various sets of cuts.  This table presents many
results in compressed form and is organized as we now explain.

Columns labeled ``Fixed center, fixed radius'' correspond to case 1 in
\S\ref{sec:galaxy_counts}--\ref{sec:flux_maps}: we simply compute the total
number (or median flux) of galaxies inside the WMAP cold spot, and report the
deviation (in ``sigmas'') from the expected value.  A positive sign indicates
a result (either count or flux) which is larger than expected; a negative sign
indicates a result which is less than the expected value.

Columns in Table~\ref{tab:cuts} labelled ``floating center, floating radius''
correspond to case 3 in \S\ref{sec:galaxy_counts}--\ref{sec:flux_maps}.
Each entry in these columns corresponds to a complete analysis along the lines
of \S\ref{ssec:case3} and has been calculated as follows.

We first compute the statistic $N_\sigma(P_0,r_0)$, defined in
Eq.~(\ref{eq:defnsigma}) to measure the ``number of sigmas'' of the worst underdense subdisc
of the WMAP cold spot $(P_0,r_0)$.
As explained in \S\ref{ssec:case3}, this statistic cannot be used directly to assess
significance since the subdisc is an {\em a posteriori} choice.
We therefore define
\be
{\mathcal N}_{\rm under} = \frac{N_\sigma(P_0,r_0) - \langle N_\sigma \rangle}{\Delta N_\sigma}  \label{eq:nunder_def}
\ee
where $\langle N_\sigma \rangle$ and $(\Delta N_\sigma)$ are the mean and RMS of the quantity
$N_\sigma(P_0,r_0)$ taken over an ensemble of regions with the same size and declination as the cold spot.
For example, with default cuts, the analysis in \S\ref{ssec:case3} can be summarized by the statement
that $N_\sigma(P_0,r_0)=-3.06$ and ${\mathcal N}_{\rm under} = (-3.06 + 2.65)/0.45 = -0.9$.
The interpretation is that the ``floating center, floating radius'' analysis has found an underdense subdisc
of the cold spot with significance 0.9$\sigma$.

The sign convention in Eq.~(\ref{eq:nunder_def}) has been chosen so that a negative
value of ${\mathcal N}_{\rm under}$ corresponds to an underdensity which is more
anomalous than the ensemble mean.
A positive sign would mean that the most underdense subdisc in the WMAP cold
spot is less anomalous than expected, when compared to an ensemble of regions with the same size
and declination as the cold spot.

We define a quantity ${\mathcal N}_{\rm over}$ in an analogous way, choosing the sign so that
a positive value corresponds to an overdense region which is more anomalous than the ensemble mean.

In Table~\ref{tab:cuts}, we report either ${\mathcal N}_{\rm under}$ (indicated by a negative sign) or
${\mathcal N}_{\rm over}$ (indicated by a postive sign), {\em whichever is more anomalous}.  
A `---' entry means that ${\mathcal N}_{\rm under}$ is positive and ${\mathcal N}_{\rm over}$ is
negative, i.e. we find no subdisc (either underdense or overdense) of the cold spot which 
is more anomalous than the ensemble mean, for a given set of cuts.

For example, with default cuts, we find a negative 
value for ${\mathcal N}_{\rm over}$, i.e. the most anomalously overdense subdisc of the WMAP cold
spot is actually less overdense than the ensemble mean.
Therefore, the corresponding entry in Tab.~\ref{tab:cuts} is given by ${\mathcal N}_{\rm under} = -0.9$.
This value summarizes the analysis from \S\ref{ssec:case3}: with default cuts, we find a subdisc which
is anomalously underdense at 0.9$\sigma$ (and any overdense subdisc is less anomalous than this).

Finally, columns in Table~\ref{tab:cuts} labelled ``fixed center, floating radius'' correspond to case 2 from
\S\ref{sec:galaxy_counts}--\ref{sec:flux_maps}, with one difference: when
reporting the significance of the most anomalous radius $r$, we incorporate
the {\em a posteriori} choice of radius by maximizing over $r$ as in case
3.  (We omitted this step for simplicity in
\S\ref{sec:galaxy_counts}--\ref{sec:flux_maps} because with our default cuts, all
choices of $r$ turned out to give very typical values.)

There are a few values in Table~\ref{tab:cuts} which might be interpreted as
statistically significant, e.g. the example which motivated this section: for
the flux range $S\ge 5$ mJy, there is a subdisc (the ``floating center, floating
radius'' case) which has galaxy counts low at 2.5$\sigma$ even after accounting for
the {\em a posteriori} choice of center and radius.  However, we note that the
statistical significance goes away when complex sources are dropped, or if we
restrict the flux range further.  Furthermore, one can find another value in
the table which supports an {\em overdensity} with the same statistical significance
(the ``fixed center, fixed radius'' galaxy count case with flux range $5\le S\le 12$ mJy).
Given the
large number of entries in Table~\ref{tab:cuts}, a few high-significance values
such as these are expected as statistical events.\footnote{To make this more
  quantitative, given only 4 independent events, the likelihood that one event
  is anomalous at the $2.5\sigma$ level is $\approx 2\sigma$, so getting a few
  such anomalous values in an analysis such as Table~\ref{tab:cuts} with many
  different choices of cuts is not surprising.}  

Analogously, for the flux analysis, there is a 2.9$\sigma$ low-flux subdisc 
in the flux range $3\le S\le 5$ mJy, but if complex sources are dropped, we
find a 2.5$\sigma$ {\em high-flux} disc in the same flux range.  
Since there is no clear
pattern to the few high-significance values, and since a high source density/flux
region is supported as well as a low source/flux region,
our interpretation of Table~\ref{tab:cuts} is that there
is no evidence for either NVSS number counts or median source fluxes which are
atypical in the WMAP cold spot.

\section{Discussion and Conclusions}
\label{sec:conclude}

In this paper, we have revisited claims from \citet{Rudnick} that there is a
cold spot in the NVSS radio survey which is statistically significant and
aligned with the cold spot found in WMAP \citep{Vielva2004,Cruz2005}.  We
found no evidence for either an underdensity in NVSS number counts, or a
region of atypical median source flux.

Our analysis incorporates systematic declination-dependent striping in NVSS,
by estimating quantities such as $\langle\Ngal(P,r)\rangle$ directly from the
data via isolatitude averaging (e.g. Eq.~(\ref{eq:nexp2})).  In an analogous
way, we incorporate statistical clustering of galaxies in NVSS by estimating
variances over isolatitude rings (e.g. Eq.~(\ref{eq:delta2})).

Simple, direct statistical tests, such as counting the total number of all
NVSS galaxies in the WMAP cold spot, or taking the median flux of all such
galaxies, do not show any statistically significant anomaly.  (This
corresponds to case 1 in \S\ref{sec:galaxy_counts}--\ref{sec:flux_maps}.)
Things only become murky when one considers statistics with many {\em a posteriori}
choices, such an anomalous subdisc of the cold spot (case 3), or a choice
of selection cuts which maximizes the quoted significance (Table~\ref{tab:cuts}).  We
have exhaustively studied many such statistical tests and argued that when the
{\em a posteriori} choices are included in the assessment of statistical
significance, there is no evidence for an NVSS ``cold spot''.

As a concrete example, consider the ``$S\ge 5$ mJy'' case in Fig.~5 of
\cite{Rudnick}, which seems to show a $\approx 5\sigma$ underdense region, if
the errorbars are taken at face value.  We agree with the number counts that
are plotted in this figure, but disagree that there is statistically
significant evidence for an underdensity.  Let us illustrate this by following
this example through the steps of this paper one at a time.  First, if we use
our most accurate prescriptions for the RMS error $(\Delta\Ngal)$ and the
expected count $\langle\Ngal\rangle$
(Eqs.~(\ref{eq:nexp2}),~(\ref{eq:delta2})), then we find that the statistical significance
drops to 3.4$\sigma$; however this ignores the effect of {\em a posteriori}
choices.  The center of this underdense region is not the WMAP cold spot
center, and if we account for this choice (and the {\em a posteriori} choice
of radius) using the method of \S\ref{ssec:case3}, then we find that the significance
decreases to 2.5$\sigma$.

This now accounts for the {\em a posteriori} choice of subdisc, and appears to
give a statistically significant result, but we have still made an {\em a
  posteriori} choice of selection cuts, by allowing complex structure and
considering only sources with flux $S\ge 5$ mJy.  When viewed in the larger
context of Table~\ref{tab:cuts}, it is seen that these choices maximize the
quoted ``number of sigmas'' of an underdensity, and simply reflect the large
number of possible choices of selection cuts: the significance goes away if
the cuts are changed slightly, and in fact a different choice of cuts would
favor an {\em overdensity} rather than an underdensity, with roughly the same
significance.  This last observation is perhaps the most convincing sign that
the apparent $\approx 2.5\sigma$ underdensity, for a single choice of
selection cuts, is spurious. 

For the median flux analysis (\S\ref{sec:flux_maps}), our conclusions are the
same: we find no evidence for atypical source fluxes in the WMAP cold spot,
after accounting for {\em a posteriori} choices.  There are a few choices of
cuts which appear to show anomalous values (if the {\em a posteriori} choice
of cuts is ignored), but the number of such values is consistent with
statistics, and the cuts can be tuned to support either a region with high or
low source density/flux with roughly the same statistical significance
(Table~\ref{tab:cuts}).

We do not see reason to give preferential treatment to {\em a posteriori} choices in
the analysis and selection cuts given in \citet{Rudnick}, and we instead
considered a range of analyzed quantities and selection criteria.
Had there been a
physical reason or a survey-specific requirement for the particular treatment
of raw data used in \citet{Rudnick}, we would have agreed with that choice
being well motivated or even necessary. However, since we do not see such
motivation, we insist on calling all such choices {\em a posteriori}.

In \cite{McEwen:2006my}, the cold spot was identified as one of 18 regions
which are ``peaks'' in the ISW cross-correlation between WMAP and NVSS.
However, this analysis was performed using wavelet smoothing with scale 250'
and would be blind to the $1^\circ$ underdense region studied in \cite{Rudnick}.
Furthermore it is not clear from the analysis in \cite{McEwen:2006my} whether
the WMAP-NVSS cross-correlation is statistically significant when restricted
to the cold spot alone.

Despite the null result of this paper, one should not be disheartened. More
detailed observation of the cold spot region in galaxy surveys will
likely improve confidence about the existence of any over/underdensity or lack
thereof. More generally, new WMAP data and the eagerly expected Planck maps
expected in a few years, combined with data from a variety of galaxy surveys
from ground and space, will provide a gold mine to search for signatures of
the early and late-universe physics in the large-scale structure and the CMB.

\section*{Acknowledgments}
We would like to thank
Lawrence Rudnick, 
Cora Dvorkin,
Eiichiro Komatsu,
Wayne Hu,
and David Spergel
for useful discussions.
KMS was supported by an STFC Postdoctoral Fellowship.
DH was supported by the DOE OJI grant under contract DE-FG02-95ER40899, 
and NSF under contract AST-0807564.

\bibliographystyle{mn2e}
\bibliography{coldspot}

\end{document}